# Energy-momentum mapping of *d*-derived Au(111) states in a thin film

P. M. Sheverdyaeva[1], R. Requist[2,3], P. Moras[1], S. K. Mahatha[1,4], M. Papagno[1,5], L. Ferrari[6], E. Tosatti[2,4,7], and C. Carbone[1]

[1] *Istituto di Struttura della Materia, Consiglio Nazionale delle Ricerche, Strada Statale 14 Km 163.5, 34149 Trieste, Italy*
[2] *International School for Advanced Studies (SISSA), Via Bonomea 265, 34136 Trieste, Italy*
[3] *Max Planck Institute of Microstructure Physics, Weinberg 2, 06114 Halle, Germany*
[4] *International Centre for Theoretical Physics (ICTP), Strada Costiera 11, 34151 Trieste, Italy*
[5] *Dipartimento di Fisica, Università della Calabria, Via P.Bucci, Cubo 31 C, 87036 Arcavacata di Rende, Italy*
[6] *Istituto dei Sistemi Complessi, Consiglio Nazionale delle Ricerche, Via del Fosso del Cavaliere, 100, 00133 Rome, Italy*
[7] *Istituto Officina dei Materiali, Consiglio Nazionale delle Ricerche, Democritos National Simulation Center, Via Bonomea 265, 34136 Trieste, Italy*

The quantum well states of a film can be used to sample the electronic structure of the parent bulk material and determine its band parameters. We highlight the benefits of two-dimensional film band mapping, with respect to complex bulk analysis, in an angle-resolved photoemission spectroscopy study of the *5d* states of Au(111). Discrete *5d*-derived quantum well states of various orbital characters form in Au(111) films and span the width of the corresponding bulk bands. For sufficiently thick films, the dispersion of these states samples the bulk band edges, as confirmed by first-principles calculations, thus providing the positions of the critical points of bulk Au in agreement with previously determined values. In turn, this analysis identifies several *d*-like surface states and resonances with large spin-splittings that originate from the strong spin-orbit coupling of the Au 5*d* atomic levels.

# I. INTRODUCTION

Angle-resolved photoemission spectroscopy (ARPES) is a powerful experimental method, based on the photoelectric effect, for studying the electronic structure of bulk crystalline solids [1]. Within an independent electron approximation, symmetry rules and conservation laws link the kinetic energy and angular distribution of the photo-emitted electrons with the electronic bands inside the crystal. While the in-plane electron wave vector ($k_{\parallel}$) is conserved through the photoemission process, the surface potential step makes the initial state wave vector along the surface normal ($k_{\perp}$) not directly measurable [2]. The most frequently used model to obtain $k_{\perp}$ assumes free electron-like final states and a mean inner potential for the electrons inside the crystal [1]. The electronic structure of many solid systems has been determined by ARPES, and it has been especially successful in the identification of critical points [3,4].

The ARPES analysis of thin films can be exploited to map the electronic structure of the corresponding bulk crystals. In thin film systems $k_{\perp}$ acquires discrete values, which are determined, in analogy with the particle-in-a-box picture, by the film thickness (i.e. the width of the potential well). The quantization of $k_{\perp}$ gives rise to a finite number of two-dimensional quantum well (QW) states that span the width of the surface-projected bands of the parent bulk material. For sufficiently thick films, the QW state dispersion closely approaches the bulk band edges, which, therefore, become measurable by ARPES without the use of approximations [5]. In particular, this method can provide the location of critical points along several bulk directions with a single measurement.

The experimental determination of the bulk band properties by thin film analysis depends critically on the ability of ARPES to observe QW states. Several studies report the detection of *sp*-like QW states in thin metal films [5-20], while comparatively few deal with *d*-derived QW states [5, 8, 21-28]. Indeed, these states are difficult to resolve when the film thickness exceeds a few monolayers, due to the overlap of narrow *d*-bands with different symmetry. Nevertheless, the $k_{\parallel}$-dependent dispersion of *d*-like QW states has been recently determined by ARPES on relatively thick Co [29] and Fe [30] films.

In the present study *d*-derived QW states of various orbital symmetries are detected by ARPES in Au(111) films grown on Ag(111). On the basis of the very good agreement between experiment and density functional theory (DFT) calculations for a free-standing Au slab, the surface-projected Au band edges as a function of $k_{\parallel}$ are determined. This information provides the location of many Au critical points, which comprise previous observations for Au bulk crystals. The identification of Au band edges in the *5d* region reveals several surface states and resonances, so far only predicted by theory, whose large spin-splittings derive from the strong spin-orbit interaction of the original *5d* atomic levels.

## II. EXPERIMENTAL AND COMPUTATIONAL METHODS

The experiments were performed at the VUV-Photoemission beamline of the Elettra synchrotron (Italy). The Ag(111) single crystal was prepared by cycles of Ar ion sputtering and annealing to 800 K. Au films were grown on the Ag substrate at 150 K. Post-deposition annealing to room temperature produces Au(111) films of uniform thickness, where *d*-like QW states could be detected by ARPES. Both substrate and films showed sharp (1×1) low energy electron diffraction patterns. Photoemission spectra were measured at 150 K, using a Scienta R-4000 electron analyzer and photon energies between 35 and 80 eV. The relatively large film thickness used in the experiment and the absence of Ag core level lines (not shown) allow us to interpret all observed features as Au-derived states. Energy and angular resolution were set at 25 meV and 0.3°, respectively.

DFT calculations were performed with QuantumEspresso [31], a plane wave pseudopotential electronic structure code. The local density approximation (LDA) in the parameterization of Perdew and Zunger [32] was used for the exchange-correlation potential. The lattice constant of the free-standing 24 monolayer Au slab was set according to the bulk experimental lattice constant (4.078Å), and the interlayer spacing within the slab was held fixed, since relaxation gave an incorrect description of the *d* band width. In calculations of the Au/Ag bilayer, the lattice constant, interlayer spacing, and atomic positions were fixed in accordance with those of bulk Au to facilitate comparison with the free-standing slab calculations. The resulting errors are negligible, since the lattice constants of Ag (4.085Å) and Au are extremely close. All calculations used ultrasoft pseudopotentials with plane wave cut-offs of 30 Ry for the wave function and 360 Ry for the charge density. Integrations over the two-dimensional Brillouin zone were performed on a 12·12 *k*-point mesh for the free-standing Au slab and a 20·20 *k*-point mesh for the Au/Ag bilayer with a smearing width of 0.020 Ry. The LDA+U method with U=1.5 eV was used [33]; this value of the Hubbard parameter was chosen to shift the center of gravity of the *d* bands down by 0.55 eV in accordance with the photoemission data.

## III. RESULTS AND DISCUSSION

Fig. 1(a) shows ARPES data along the $\bar{\Gamma}$–$\bar{K}$ direction for a 24 monolayer (ML = 2.354 Å) Au(111) film. The observed spectroscopic features are identified as the Shockley surface state near the Fermi level ($E_F$) at $\bar{\Gamma}$, sp-like bands for binding energy above 2 eV and below 6 eV and d-like bands between 2 and 6 eV. A closer look at the data reveals a manifold of d-derived states that are not present in bulk Au(111). As an example, Fig. 1(b) reports energy distribution curves extracted from the area enclosed by the white box in Fig. 1(a), where five dispersive peaks are highlighted by vertical ticks. This is the signature of d-derived QW state formation, which can be identified by ARPES only in atomically uniform films, due to the narrow d-band width [21]. Fig. 1(c-f) reports ARPES spectra measured in the same energy-momentum region at different photon energies. These data are displayed as second derivatives along the energy axis to enhance the sensitivity to low-intensity features. For all photon energies the electronic states are detected at the same position. The absence of $k_\perp$-dependence confirms the two-dimensional character of the Au film electronic structure. Matrix element effects cause strong intensity variations of the photoemission signal as a function of photon energy and experimental geometry.

DFT calculations for a free-standing 24 ML Au(111) film with the spin-orbit interaction included have been performed to identify the spectroscopic features in the ARPES data (Fig. 2(a)). The color code in the figure refers to the spatial character of the electronic states. Surface states are indicated by red lines, whose thickness is proportional to the charge density in the two topmost layers. QW states delocalized over the whole film are represented by blue lines. The grey shaded area shows the surface-projected bulk Au bands. Notably, at the selected film thickness (24 ML) the d-like QW states densely sample the corresponding bulk bands, thanks to their small energy separation. In particular, the distance between the surface-projected bulk band edges and the nearest QW states is less than 60 meV for all $k_\parallel$ values. Therefore, the width of the bulk bands as a function of $k_\parallel$ can be determined by the analysis of the QW states of corresponding symmetry with an accuracy better than 120 meV.

Fig. 2(b-d) reports theoretical Au bulk bands (labeled B0-B5) along the $\Gamma L$, $LX$ and $A_1 A_2$ lines [34, 35], in order to link film to bulk properties. These lines project onto the $\bar{\Gamma}$, $\bar{M}$ and $\bar{K}$ points, according to the scheme reported on the right hand side of Fig. 2(a). The bottom of B0 and the top of B5 have prevalently sp-character, while the other bands have prevalently d-like character. Red dots and lines indicate the experimental critical points, determined according to the procedure explained below. The connection between QW states and bulk bands is further described in the Supplemental Material [36].

**A. Quantum well states**

In this section we show how the critical points of bulk Au are experimentally determined by thin film analysis. To this end, Fig. 3 presents the second derivative of the ARPES spectra for the 24 ML Au film, without (left column) and with (right column) theoretical bands overlaid. Excellent agreement between experiment and theory is found once a rigid downward shift of 0.27 eV is applied to the calculated bands. Based on this agreement, we distinguish in the experimental dataset QW states of different orbital character and determine the related bulk band edges as a function of $k_\parallel$. The energy of these edges at $\bar{\Gamma}$ and $\bar{M}$ defines the experimental bulk critical points at $\Gamma$, L and X [36]. A few examples of this procedure are discussed in the following.

Fig. 3(a,b) displays data taken at 65 eV photon energy along the $\bar{K}–\bar{\Gamma}–\bar{K}$ direction. The nearly parabolic B5-derived QW states at the top of the image overlap with flatter B4-derived QW states at larger binding energies. The topmost edge of the B4 states is marked by a high intensity surface resonance. Steeply downward dispersive features are associated to the B3-bulk band. Finally, an unresolved bundle of B5-derived QW states with negative aperture is seen at the bottom of the image. This analysis identifies the band edges of different QW state families as a function of $k_\parallel$. In particular, the position of these edges at $\bar{\Gamma}$ coincides with the critical points of bulk Au along the $\Gamma L$ direction. The intense band at 2.28 eV is the top of the B4-derived states and corresponds to the $L_{5,6}^+$ critical point. The bright feature at the bottom of the image (3.92 eV) is connected to the global minimum of B5 between $\Gamma$ and L (see Supplemental Material [36]). From ARPES data collected over an extended energy range and with a few different photon energies, several other critical points are found along $\Gamma L$ (red dots in Fig. 2(b)).

Analogously to Fig. 3(a,b), different families of QW states can be observed in the other panels of Fig. 3. Fig. 3(c,d) shows data along the $\bar{K}–\bar{M}–\bar{K}$ direction in the proximity of $\bar{M}$. B4-derived QW states are identified at the top of the panel and an overlapping between B3- and B2-derived QW states at the bottom of it. The dispersion of these bands in the perpendicular direction (i.e. $\bar{\Gamma}–\bar{M}–\bar{\Gamma}$) can be followed in Fig. 3(e,f). Again, the comparison between experiment and theory identifies the surface-projected band edges and the location of several critical points, which are reported as red dots along the LX direction (Fig. 2(c)).

Table 1 shows the very good agreement between Au bulk critical points, previously determined by ARPES [37-45], and the results of the present investigation. Notably, Au bulk data have been obtained from different crystal faces by systematic change of the photon energy to scan the high symmetry direction perpendicular to the surface [37-43] or by complex triangulation methods [44,45]. Instead, our method gives the location of the same bulk critical points by the analysis of a single Au film with a few different photon energies.

In Fig. 3(g,h) we attempt an examination of the bands along the $A_1A_2$ line, which projects on the $\bar{K}$ point. In analogy to the previous cases, QW states of different orbital origin are resolved. The narrow width and complex dispersion of the bands at $\bar{K}$ prevent a precise location of the corresponding bulk band edges along $A_1A_2$ . For this reason, dashed lines, rather than dots, are used in Fig. 2(d).

Table 1. Experimental critical points for Au(111). The error bar is defined as the full width at half maximum of the related photoemission peak.

| Symmetry label | Energy, eV | |
| --- | --- | --- |
| | This work | Previous works |
| $\Gamma_8^+$ | 5.90±0.12 | 6.00[37], 6.01 [38] |
| $\Gamma_7^+$ | 4.60±0.10 | 4.60[37], 4.68 [38] |
| $\Gamma_8^+$ | 3.62±0.08 | 3.65[37], 3.71 [38] |
| $L_{5,6}^+$ | 6.22±0.14 | 6.23 [37], 6.20 [38] |
| $L_4^+$ | 5.00±0.12 | 4.88 [37], 5.00 [38] |
| $L_4^+$ | 3.04±0.06 | 3.20 [38] |
| $L_{5,6}^+$ | 2.28±0.07 | 2.30 [38] |
| $X_7^+$ | 1.60±0.04 | 1.60 [39], 1.90 [40] |
| $X_6^+$ | 2.76±0.06 | 2.40 [40] |
| $X_7^+$ | 3.04±0.06 | 3.00 [40] |
| $X_7^+$ | 7.78±0.16 | 7.50 [40] |

**B. Effects of the hybridization with the Ag substrate**

Photoemission studies of Au films on Ag(111) [20, 46, 47] have revealed the influence of the substrate on the Au *sp*-derived QW states. Hybridization between Ag and Au electrons gives rise to strongly reduced interface reflectivity and modifications of the Au QW state dispersion with respect to the case of a free-standing Au(111) film. At variance with *sp*-states, Au *d*-states are much more localized and have much narrower bandwidths. In general, these properties tend to increase the interface reflectivity for *d*-levels, so that the particle-in-a-box picture inherent in our free-standing Au film calculations appears to be a good first approximation to the behavior of *d*-like QW states in the Au/Ag system.

In order to quantify the effects of hybridization with the Ag substrate, we performed calculations for a 24 ML Au film on a 6 ML Ag substrate. The resulting band structure is shown in Fig. 4. The spatial localization of the states is represented by coloring the lines according to the total weight of the state on the Au film; blue represents states prevalently localized in Au, while the continuous progression from blue to green to yellow indicates increasing delocalization into the Ag substrate. To improve the clarity of the plot, the number of overlapping bands is reduced by applying a cutoff to omit the states which are most strongly localized on Ag (weight on Au < 0.35). The grey shaded regions display the surface-projected bulk bands of Au, similarly to Fig. 2(a). In agreement with previous studies, the B5-derived *sp*-like states delocalize significantly into the Ag substrate as shown by the green color of the lines. There are some significant displacements of the *sp*-derived QW states with respect to the corresponding results for a free-standing Au slab. On the other hand, hybridization-induced distortions are smaller for the *d*-derived states and observed only far away from the surface-projected bulk band edges (see Supplemental Material [36]). This proves the validity of the analysis reported in section III A, which is based on free-standing film calculations.

From the experimental point of view, the *d*-like QW states in the Au film are very well defined for binding energies less than about 4 eV, where coupling with Ag states is negligible. For deeper energies, the combined effects of lifetime broadening and hybridization could explain the blurring of the electronic states in the ARPES signal (see Fig. 5). Nevertheless, Au-related *d*-like features are found near all the band edges, as predicted by the calculations shown in Figs. 2 and 4.

**C. Surface states**

The analysis reported in section III A determines the location of the surface-projected band gaps of bulk Au(111) and allows the identification of several *d*-like surface states and resonances. Theory predicts large spin-splittings for most of them, due to the spin-orbit coupling of the *5d* atomic levels. We use the labeling of the surface states according to Refs. [34,48].

Three surface states detected along the $\overline{\Gamma}$–$\overline{M}$ direction (Fig. 5(a,b)) have already been reported in the literature. The L-gap Shockley surface state, near $E_F$ at $\overline{\Gamma}$, and the S2 state, laying 8 eV below it, have similar Rashba-Bychkov-type spin-splitting [20,34], which is not resolved in the present dataset. The properties of the S8 surface state have been described in detail in our previous work [49]. All other surface features along $\overline{\Gamma}$–$\overline{M}$ are resonances. We notice that the S14 state is the analog of the Tamm surface state at the $\overline{M}$ point of Cu(111) [50], but turns out to be a surface resonance in Au(111), as already pointed out in Ref. [34].

We focus now on the $\overline{K}$ point, where theory predicts a number of surface states that have not been experimentally detected so far. Fig. 6(a,b) shows the highest $d$ band gap along $\overline{\Gamma}$–$\overline{K}$–$\overline{M}$. The S3a,b states, with prevalently $d_{xz}$, $d_{yz}$ character, display a nearly constant spin-splitting as a function of $k_{\parallel}$ (0.30 eV at $\overline{K}$). Analogous states for Ir(111) have been discussed in Ref. [51]. Two additional states detected in the same gap are unaccounted for by DFT calculations. The flat feature located at 2.9 eV displays an abrupt attenuation as $k_{\parallel}$ moves away from $\overline{K}$. This behavior could be the fingerprint of a surface state to resonance transition upon coupling to bulk bands of the Ag substrate. A second flat band is observed at 3.05 eV. Trial tight binding calculations based on DFT parameters suggest that these two features are split-off surface states, generated near the bottom of the B4 band by suitable modifications of the surface potential.

Four surface states are located in the band gap displayed in Fig. 6(c,d). S4a-c are very well reproduced if calculations are shifted by 0.18 eV to higher binding energies. S4b and S4c form a spin-orbit split pair of surface states, with a maximum splitting of about 1 eV at $k_{\parallel} = 0.7$ Å$^{-1}$. They belong to different irreducible representations [34] and are expected to open a gap of 0.04 eV at $\overline{K}$, which is not resolved in the experimental dataset. S4d lies very close to the B2 bulk band edge. Part of it is probably observed around $k_{\parallel} = 0.75$ Å$^{-1}$, where an avoided crossing with the S4c state is expected.

Finally, in the energy-momentum region displayed in Fig. 6(e,f) we observe the S5 family of states. While the dispersion of S5a is clearly identified, S5b and S5c are too close in energy to be individually resolved near $\overline{K}$.

Table 2 compares experimental and calculated binding energies for several surface states and resonances at the high symmetry points of the surface Brillouin zone. The LDA+U method used in the present work improves the agreement with the experimental data, with respect to the LDA results of Ref. [34]. However, significant discrepancies remain, as one can see from the last column of the table. The difference between theory and experiment increases from $E_F$ to approximately 2 eV binding energy, where it reaches 0.42 eV and starts to decrease, becoming very little for binding energies higher than 4.5 eV. These discrepancies cannot be ascribed to hybridization with the Ag substrate (see Supplemental Material [36]), due to the highly localized character of the $d$-like surface states. Instead, we attribute them to an overestimation of the $d$ band width in the LDA+U method.

Table 2. Binding energy (eV) of *d*-like surface states and resonances of Au(111). Labels are given in accordance with Refs. [34, 48]. The last column reports the difference between theoretical and experimental values in the present work.

| Surface feature | Symmetry point | Theoretical values [34] | Present work | | |
|---|---|---|---|---|---|
| | | | Theoretical values | Experimental values | difference |
| S14 | $\bar{M}$ | – | 1.57 | 1.70 | -0.13 |
| S13 | $\bar{\Gamma}$ | – | 1.98 | 2.40 | -0.42 |
| S11 | $\bar{\Gamma}$ | – | 2.00 | 2.40 | -0.40 |
| S12 | $\bar{\Gamma}$ | – | 2.70 | 3.10 | -0.40 |
| S3a | $\bar{K}$ | 2.9 | 3.26 | 3.58 | -0.32 |
| S3b | $\bar{K}$ | 3.2 | 3.56 | 3.78 | -0.22 |
| S4a | $\bar{K}$ | 3.7 | 4.10 | 4.30 | -0.20 |
| S1a | $\bar{\Gamma}$ | 3.7 | 4.20 | 4.24 | -0.04 |
| S4b,c | $\bar{K}$ | 4.0 | 4.36,4.39 | 4.56 | -0.20 |
| S9 | $\bar{M}$ | – | 4.75 | 4.82 | -0.07 |
| S4d | $\bar{K}$ | 4.7 | 5.04 | – | – |
| S1b | $\bar{\Gamma}$ | 5.0 | 5.60 | 5.62 | -0.02 |
| S8 | $\bar{M}$ | 5.4 | 5.90 | 5.93 | -0.03 |
| S5a | $\bar{K}$ | 5.5 | 5.93 | 5.92 | 0.01 |
| S5b | $\bar{K}$ | 5.7 | 6.15 | 6.12 | 0.03 |
| S5c | $\bar{K}$ | – | 6.23 | – | – |
| S2 | $\bar{\Gamma}$ | 7.6 | 7.85 | 7.80 | 0.05 |
| S7 | $\bar{M}$ | 6.6 | 7.10 | 7.02 | 0.08 |

**IV. Conclusions**

We have reported a detailed investigation of the electronic structure of a 24 ML Au(111) film. Despite the relatively large film thickness, ARPES is able to resolve QW states of *d* character, even in regions where two or more groups of QW states overlap. DFT calculations reproduce the experimental observations very well. Based on this agreement, the locations of several critical points of Au have been determined and found to be very similar to existing data for bulk Au. These results demonstrate that ARPES thin film analysis is a simple and efficient method for studying the surface-projected band structure of bulk materials.

Additionally, we have experimentally identified most of the *d*-like surface states and resonances predicted by DFT calculations, thanks to the precise determination of the Au band edges. A number of surface features, located at $\overline{K}$ and showing large spin-orbit splittings, have been observed for the first time.

**Acknowledgements**

We acknowledge the "Progetto Premiale, Materiali e disposivi magnetici e superconduttivi per sensoristica e ICT" of the Italian Ministry of Education, University and Research (MIUR), Italy. Work at SISSA was supported by PRIN-COFIN contract 2010LLKJBX 004, and in part by ERC Advanced Grant No.\ 320796-MODPHYSFRICT, by the Swiss National Science Foundation through a SINERGIA contract CRSII2\_136287, and by COST Action MP1303.

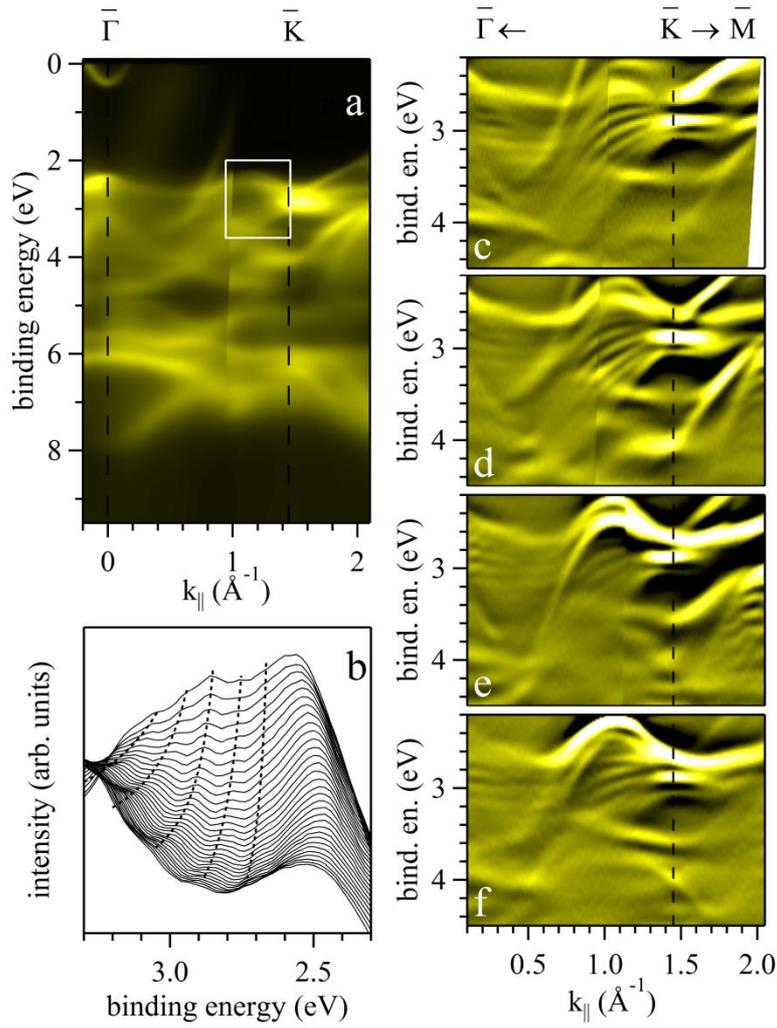

**Figure 1**. (a) ARPES spectra for a 24 ML Au film on Ag(111) measured along the $\bar{\Gamma}-\bar{K}$ direction at 45 eV photon energy. (b) Energy distribution curves relative to the area enclosed in a white box in panel (a). Vertical ticks mark the band dispersion of five B4-derived QW states. (c-f) Second derivative ARPES spectra of the same energy-momentum region acquired with photon energies of (c) 35, (d) 45, (e) 65 and (f) 80 eV.

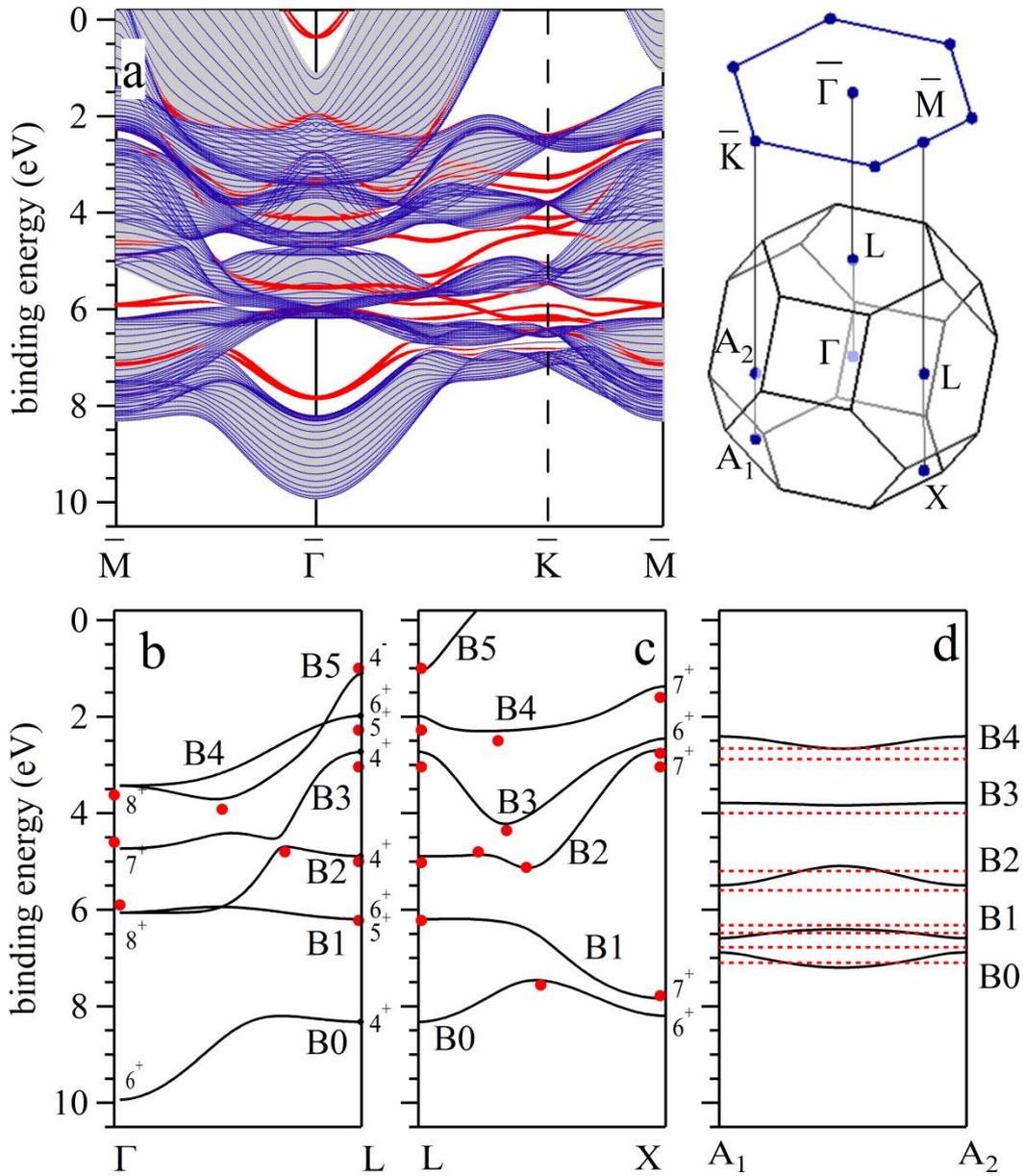

**Figure 2**. (a) DFT calculations for a free-standing 24 ML slab of Au(111) along the $\overline{M}$–$\overline{\Gamma}$–$\overline{K}$–$\overline{M}$ direction. The grey shaded region indicates the surface-projected Au bulk states. (b-d) Bulk bands of Au(111) along (b) ΓL, (c) LX and (d) $A_1A_2$ lines [35]. Red dots indicate the experimental critical points. The irreducible representations at Γ, L and X are reported [34].

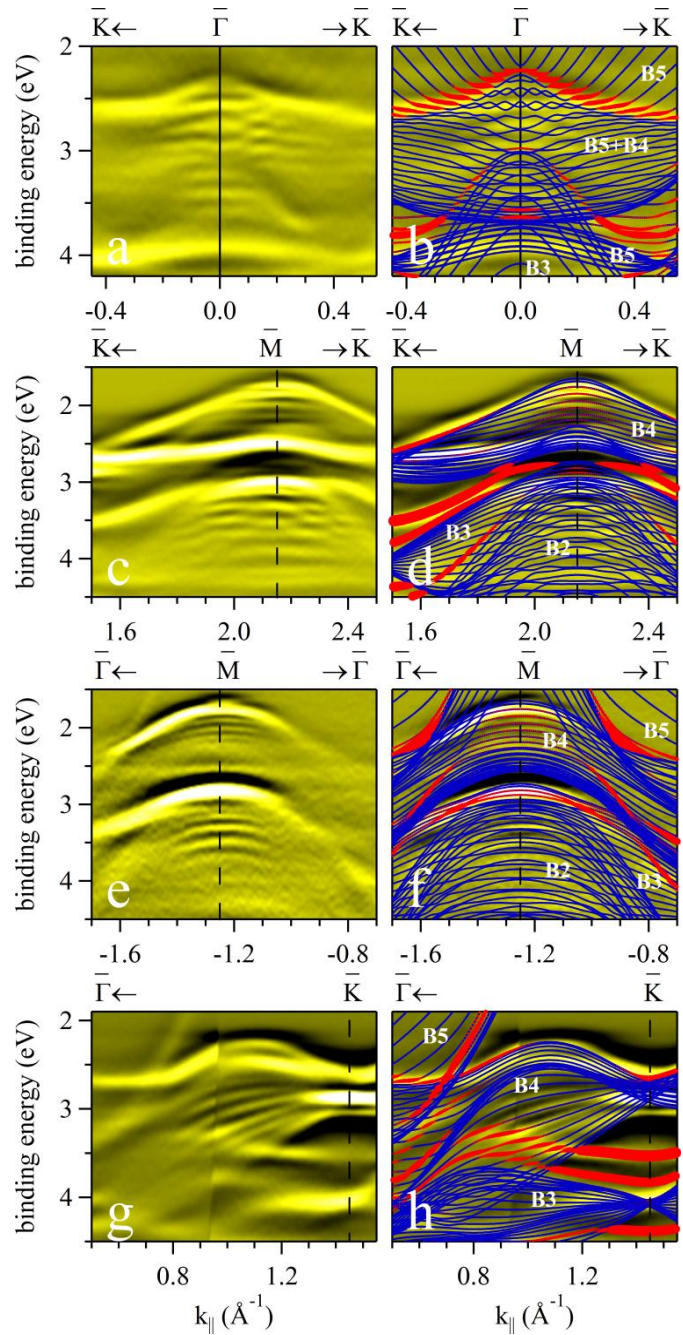

**Figure 3**. Second derivative photoemission spectra and corresponding DFT calculations (shifted down by 0.27 eV) for the 24 ML Au(111) film. (a,b) $\overline{K}$–$\overline{\Gamma}$–$\overline{K}$ direction, 65 eV photon energy. (c,d) $\overline{K}$–$\overline{M}$–$\overline{K}$ direction, 65 eV photon energy. (e,f) $\overline{\Gamma}$–$\overline{M}$–$\overline{\Gamma}$ direction, 55 eV photon energy. (g,h) $\overline{\Gamma}$–$\overline{K}$–$\overline{M}$ direction, 45 eV photon energy.

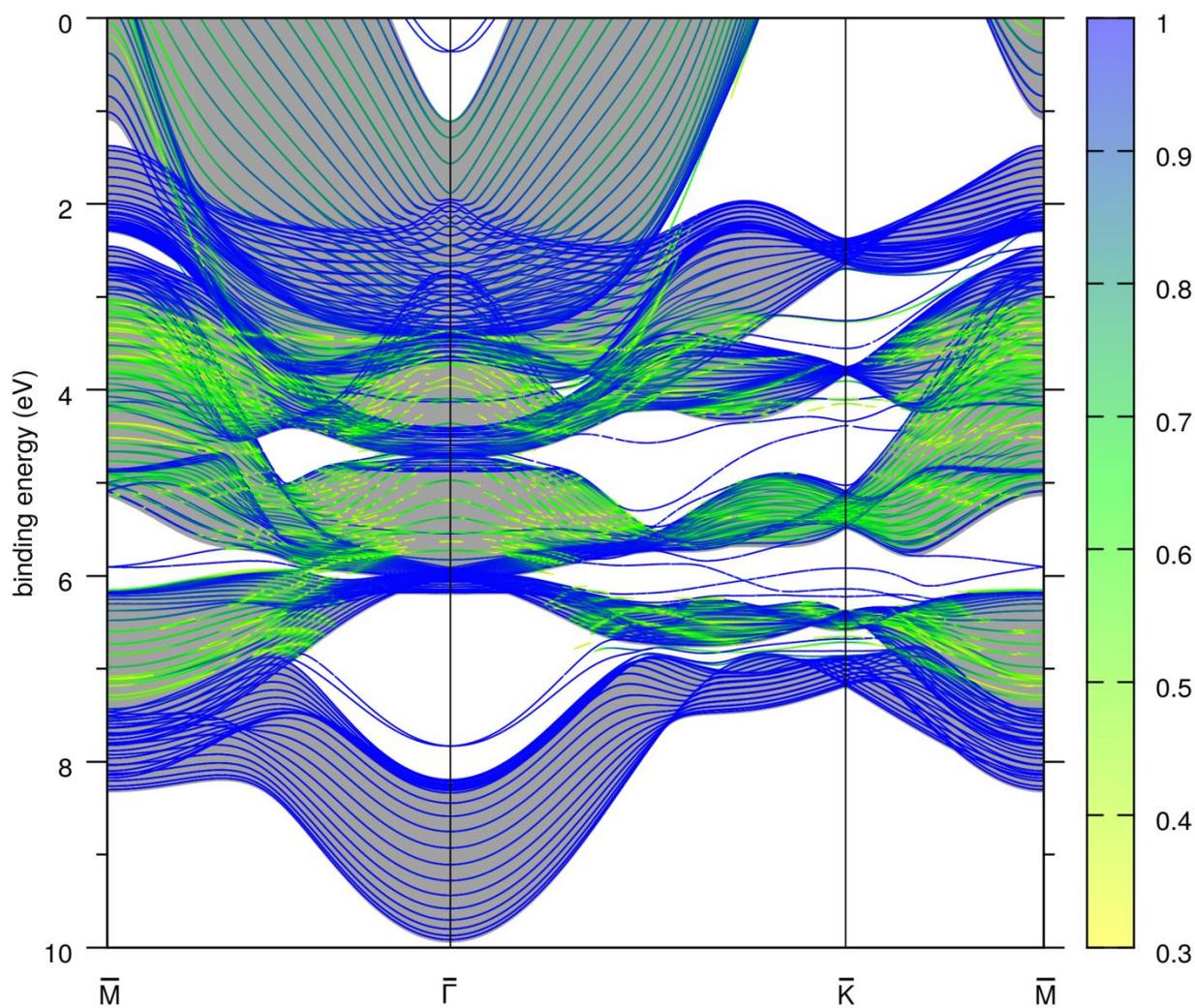

**Figure 4**. DFT band structure for a 24 ML Au(111) film on a 6 ML Ag(111) substrate. The color scale represents the spatial localization of the states inferred from their total weight on all Au layers. The gray shaded region shows the surface-projected Au bulk bands.

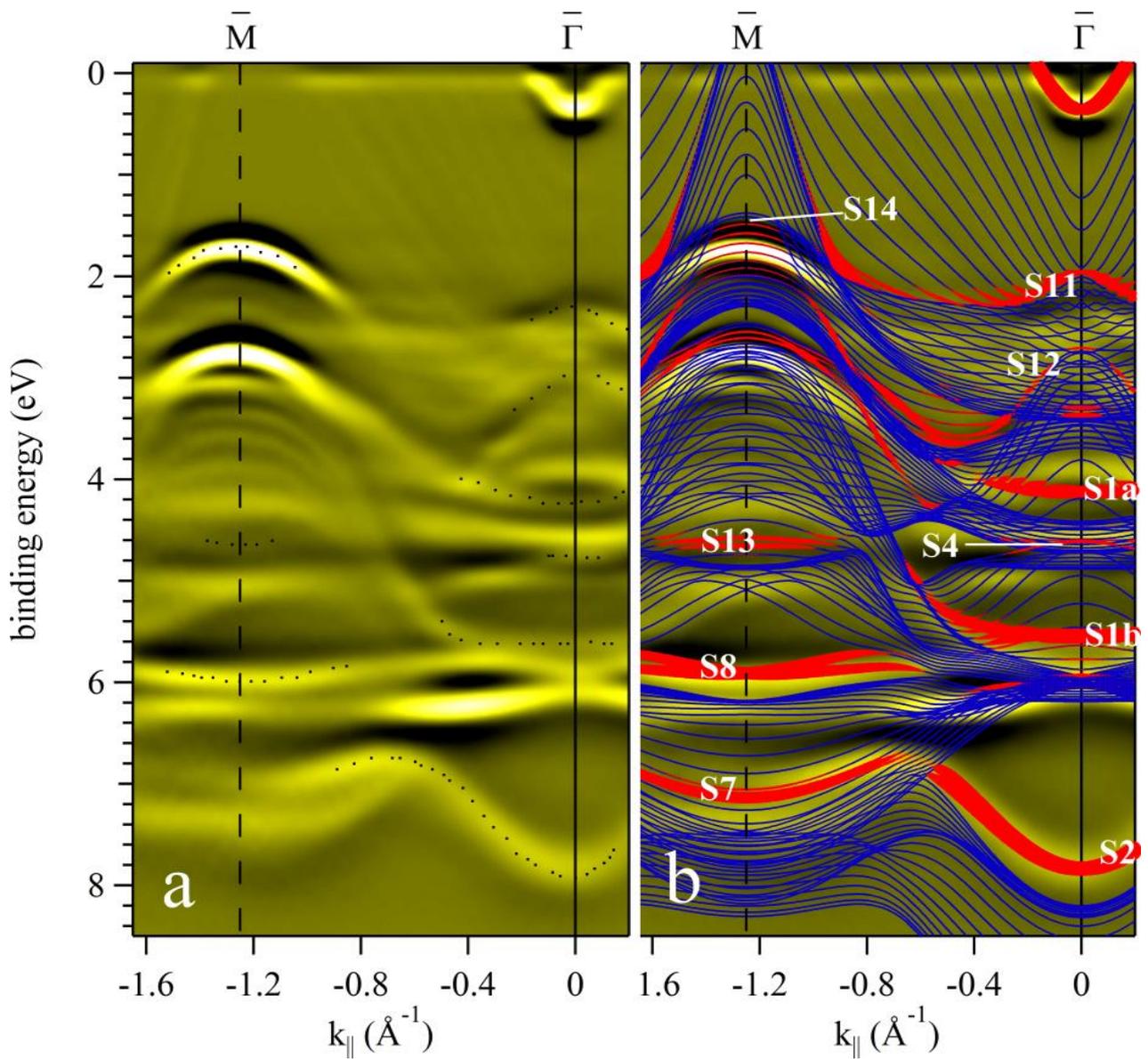

**Figure 5**. (a) Second derivative ARPES spectra at 65 eV photon energy and (b) corresponding DFT calculations for the 24 ML Au(111) film along the $\overline{\Gamma}$–$\overline{M}$ direction. No energy shift is applied to the calculated bands. Black dots mark the experimental band dispersion of surface related features.

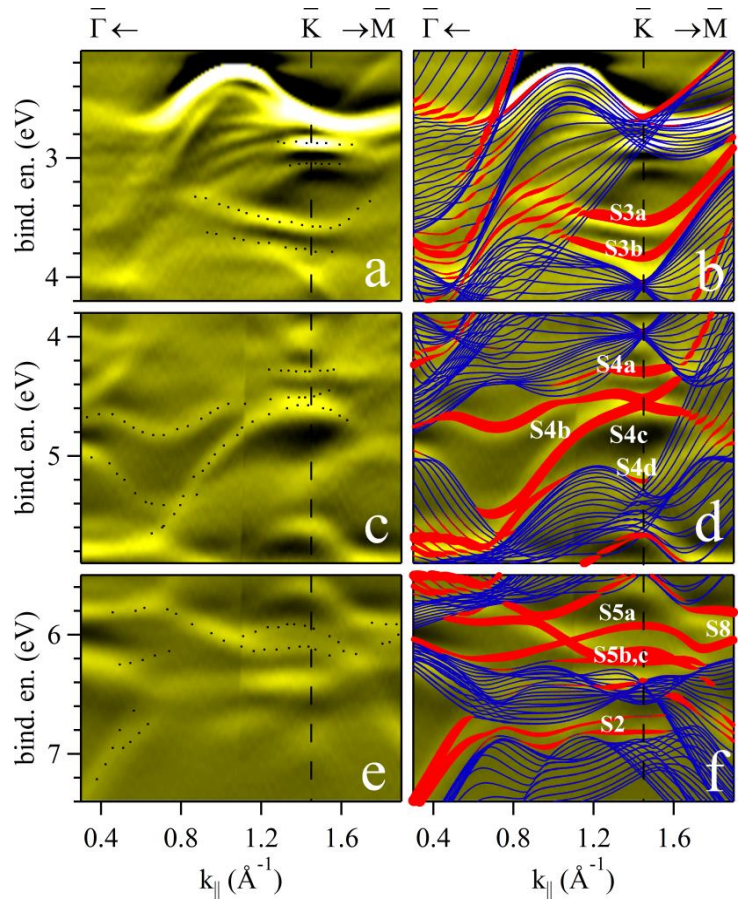

**Figure 6**. Second derivative ARPES spectra acquired along the $\overline{\Gamma}$–$\overline{K}$–$\overline{M}$ direction in the vicinity of $\overline{K}$ and corresponding DFT calculations for a 24 monolayer Au(111) film. (a,b) Photon energy 80 eV. Calculated bands are shifted downwards by 0.27 eV. (c,d) Photon energy 65 eV. Calculated bands are shifted downwards by 0.18 eV. (e,f) Photon energy 65 eV. No energy shift is applied. Black dots mark the experimental band dispersion of surface related features.